\documentclass[12pt]{article}
 \usepackage{amssymb}
\usepackage{amsfonts,amsmath,epsfig,psfrag}
\usepackage{amsmath}
\usepackage{graphicx,tikz}
 \makeatletter \oddsidemargin 0.0cm \evensidemargin 0.0cm
\newcommand{\be}{\begin{equation}}
\newcommand{\en}{\end{equation}}
\newcommand{\la}{\label}
\newcommand{\ep}{{\epsilon}}
\newcommand{\paa}{\partial}
\def\rr#1{(\ref{#1})}
\def\bm#1{\mbox{\boldmath{$#1$}}}
\def\ii{{\rm i}}

\def\citet#1{#1}
\def\citep#1{(#1)}
\def\rinf{r_{\infty}}
\def\zinf{z_{\infty}}

\title{Characterisation and stability of localised bulging/necking in inflated membrane tubes}
\author{S. P. Pearce and Y.B. Fu\\
\small{Department of Mathematics, Keele University, ST5 5BG,
U.K.}}
\date{}
\begin{document}
\renewcommand{\theequation}{\arabic{section}.\arabic{equation}}
\maketitle

\section*{Abstract}
We consider localised bulging/necking in an inflated hyperelastic membrane tube with closed ends. We first show that the initiation pressure for the onset of localised bulging is simply the limiting pressure in uniform inflation when the axial force is held fixed. We then demonstrate analytically how, as inflation continues, the initial bulge grows continually in diameter until it reaches a critical size and then propagates in both directions. The bulging solution before propagation starts is of the solitary-wave type, whereas the propagating bulging solution is of the kink-wave type. The stability, with respect to axially symmetric perturbations, of both the solitary-wave type and the kink-wave type solutions is studied by computing the Evans function using the compound matrix method. It is found that when the inflation is pressure-controlled, the Evans function has a single non-negative real root and this root tends to zero only when the initiation pressure or the propagation pressure is approached. Thus, the kink-wave type solution is probably stable but the solitary-wave type solution is definitely unstable.

\section{Introduction}
When a cylindrical membrane tube with closed ends is inflated by an internal pressure, such as via air pumping, a localised bulge forms when the pressure reaches a critical value $p_{\rm cr}$. As more air is pumped
into the tube, the pressure drops but the radius at the centre of the bulge will increase until it reaches a maximum value $r_{\rm max}$. With continued inflation, the pressure stays at a constant value $p_{\rm m}$, and the bulge spreads in both directions while the radius at the centre of the bulge maintains the maximum value $r_{\rm max}$. This process is well-known
and has been described in a number of numerical and experimental studies such as Kyriakides and Chang (1990, 1991), Shi and Moita (1996),
Pamplona {\it et al} (2006), and Goncalves {\it et al} (2008). Various aspects of this process have also been examined in many analytical studies.
The earliest analytical study seems to be that by Kydoniefs and Spencer (1969) who obtained an exact solution for an inflated membrane tube sealed by a rigid plug at each end and modeled by the Mooney-Rivlin material model. Yin (1977) proposed a method for characterising the kink-wave type bulging solution.
Stability and bifurcation of the uniformly inflated state was studied by Corneliussen and Shield (1961), Shield (1972), Haughton and Ogden (1979),
and Chen (1997). Chater and Hutchinson (1984) recognised that this process shared the same features as a family of other problems such as propagating buckles in long metal tubes under external pressure (Kyriakides, 1981), propagating necks in some polymeric materials when pulled in tension (Hutchinson and Neale 1983) and stress-induced phase transformations (Ericksen 1975). They used this connection to demonstrate that the propagating pressure $p_{\rm m}$ could be determined by the Maxwell equal-area rule.
The so-called limit-point instability, corresponding to the fact that the pressure-volume curve in uniform inflation has a turning point, was thought to
be relevant to this process, and there are a number of studies devoted to the determination of this limiting pressure including Alexander (1971),
Benedict {\it et al} (1979), and more recently Kanner and Horgan (2007). However, the exact correspondence between the limiting pressure and the initiation pressure for onset of localised bulging does not seem to have been fully appreciated; this connection is now clear if one compares Fu {\it et al}'s (2008) equation (6.2) with Chen's (1997) expression (25). We observe that in this correspondence the limiting pressure must be evaluated at fixed axial force. Thus, if the axial stretch at the two ends of a very long tube is maintained at unity, which was often assumed to simplify analysis, then the corresponding limiting pressure will
be different from the initiation pressure since a variable axial force would be required to maintain unit axial stretch at infinity. In this connection,
we note that the solid line in Kyriakides and Chang's (1990) Figure 9  seems to have been miscalculated; this figure shows unsatisfactory disagreement
between the theoretical limiting pressure and their experimental result for the initiation pressure. Fu {\it et al} (2008) recalculated the solid line
in Kyriakides and Chang's (1990) Figure 9  and found almost perfect agreement between the theoretical limiting pressure and the experimental initiation pressure.

This paper may be viewed as a sequel to our previous study, Fu {\it et al} (2008), hereafter referred to as FPL, where it was shown that the onset of localised bulging or necking
corresponds to a bifurcation at zero mode number and the mode shape can only be described by a weakly nonlinear post-bifurcation analysis.
To simplify analysis, the axial stretch at infinity was assumed to be unity. In this paper, we consider the more realistic case in which the tube
has closed ends, and our study will not be confined to the near-critical regime. Instead we aim to characterise the entire bulging or necking process,
from its weakly nonlinear initial onset to the fully nonlinear propagation stage, and to assess the stability properties
of the bifurcated solution in each stage. Our study is motivated by our belief that insights derived from the inflation problem will help with our understanding
of related problems, such as kink-band formation in fibre-reinforced composites (see, e.g., Fu and Zhang 2006), which share the same features but for which analytical results are
much harder to come by. Of course, the present study is also relevant to the continuum-mechanical modeling of aneurysm formation and growth
(Humphrey and Canham 2000, Watton {\it et al} 2004, Vorp 2007, Haughton and Merodio 2009).

The rest of this paper is divided into seven sections as follows. After formulating the problem and writing down
the governing equations in the next section, we present in Section 3 diagrams of $r(0)$ as a function of $r_\infty$ for three strain-energy
functions, where $r_\infty$ and $r(0)$ are the radii at infinity and at the centre of the bulge, respectively. We use these diagrams as the basic tool
to characterise the entire bifurcation process. In Section 4 we use the phase plane method to provide a different perspective on the bifurcation process.
In Section 5, we study the stability of the weakly nonlinear initial bulging solution with respect to axially symmetric perturbations,
with the initial bulging solution obtained using the procedure explained in FPL. We use this case to explain our method of stability analysis and to validate our
numerical results in the following section.
In Section 6, stability of the fully nonlinear bulging solution is studied with respect to axially symmetric perturbations. Results are compared with those
obtained in the previous section when specialising to the near-critical regime. The paper is concluded with a summary and additional remarks.

\section{Governing equations}
\setcounter{equation}{0}
We model the tube as an incompressible, isotropic, hyperelastic,
cylindrical membrane. The tube is assumed to have a constant undeformed
radius $R$ and a constant undeformed thickness $H$.
We shall only be concerned with localised solutions, and assume that the tube is long enough for end effects to be negligible.
Thus, such a long tube may be conveniently viewed
to be infinitely long, and we shall refer to end conditions simply as conditions imposed at infinity.
We use cylindrical polar coordinates throughout this study, and so the
undeformed configuration is given by coordinates $(R, \Theta, Z)$.

The undeformed tube is subject to a uniform internal pressure,
which drives the deformation. We assume that the axisymmetry remains
throughout the entire deformation, and hence the deformed configuration is
expressed using cylindrical polar coordinates $(r, \theta, z)$, where
$r=r(Z, t),\, \theta = \Theta, \,z=z(Z, t)$, and $t$ denotes time.

The principal directions of the deformation correspond to the lines of
latitude, the meridian and the normal to the deformed surface. Hence the
principal stretches are given by,

\begin{equation}
\lambda_{1} = \frac{r}{R}, \qquad \lambda_{2}=\sqrt{r'^2+z'^2},
\qquad \lambda_{3} = \frac{h}{H}, \label{lambdas}
\end{equation}
where the indices $1,2,3$ are used for the circumferential, axial and
radial directions respectively, a prime represents differentiation
with respect to $Z$, and $h$ denotes the deformed
thickness.

The principal Cauchy stresses $\sigma_{1}, \sigma_{2}, \sigma_{3}$ in the deformed configuration for an
incompressible material are given by
\begin{equation}
\sigma_{i} = \lambda_{i} W_{i} - p, \qquad i =1,2,3 \,\,\,\, \hbox{(no
summation)},
\end{equation}
where $W=W(\lambda_{1},\lambda_{2},\lambda_{3})$ is the strain-energy
function, $W_{i} = \paa W/\paa \lambda_{i}$, and $p$ is the pressure
associated with the constraint of incompressibility; see Ogden (1997) for further details. Utilising the
incompressibility constraint $\lambda_{1}\lambda_{2}\lambda_{3} = 1$ and
the membrane assumption of no stress through the thickness direction,
$\sigma_{3} = 0$, we find
\begin{equation}
\sigma_{i} = \lambda_{i} \hat{W}_{i}, \qquad i=1,2, \label{sigW}
\end{equation}
where $\hat{W}(\lambda_{1},\lambda_{2}) =
W(\lambda_{1},\lambda_{2},\lambda_{1}^{-1}\lambda_{2}^{-1})$ and
$\hat{W}_1=\paa \hat{W}/\paa \lambda_1$ etc (Haughton and Ogden 1979).

The equations of motion can be derived from the exact field equations of general nonlinear shell theory, e.g. Budiansky (1968), but Epstein and Johnson (2001) gave a very readable self-contained derivation. We quote their results and rewrite them in the form:
\begin{equation}
\left[ R \sigma_2 \frac{z'}{\lambda_2^2}\right]'-P^{*} \,r r'=\rho R \ddot{z},\quad \left[ R \sigma_2 \frac{r'}{\lambda_2^2}\right]'-\frac{\sigma_1}{\lambda_1}+P^{*} \,r z'=\rho R \ddot{r}, \label{mot}
\end{equation}
where $P^{*}$ is the internal pressure divided by the original wall thickness,  $\rho$ is the density of the material and a superimposed dot represents differentiation with respect to time. We note that in the static case \rr{mot} can be rewritten to give the equilibrium equations in FPL. Additionally, we non-dimensionalise the length variables with respect to the undeformed radius $R$ by setting $R=1$.

We initially look for static solutions of \rr{mot} which have uniform cross-section far away from any bulge or neck, with $r(Z) \to \rinf R,\, z(Z) \to \zinf Z$ as $Z \to \infty$, where
 here and hereafter we write $\zinf$ for $z'(\infty)$ to simplify notation. This extends the work in FPL, where the remote axial stretch $\zinf$ was set to be unity. Therefore, evaluating \rr{mot} in this uniform section we find a relation for the pressure as
\begin{equation}
P^{*}=\frac{\hat{W}_1(\rinf,\zinf)}{\rinf \zinf}, \la{pressure}
\end{equation}
which will enable us to use $\rinf$ or $\zinf$, instead of $P^{*}$, as the control parameter.

As discussed in FPL, two integrals of the equilibrium equations exist, given by,
\begin{equation}
\hat{W}-\lambda_2 \hat{W}_2= C_1=\hat{W}^{(\infty)}- \zinf \hat{W}_2^{(\infty)}, \label{equib2}
\end{equation}
\begin{equation}
\frac{\hat{W}_2 z'}{\lambda_2}-\frac{1}{2}P^{*} \lambda_1^2 R=C_2= \hat{W}_2^{(\infty)}-\frac{1}{2} P^{*} \rinf^2, \label{equib1}
\end{equation}
where a superscript $\infty$ represents evaluation at $\lambda_1 = \rinf,
\lambda_2 = \zinf$, and the conditions at infinity have been applied to determine the constants $C_1$ and $C_2$. We note that equation \rr{equib1} represents
constancy of the resultant  force   in the $Z$ direction, whereas the conservation law  \rr{equib2} was first derived by Pipkin (1968).

For an infinite tube with open ends the remote axial stretch $\zinf$ represents a prestrain of the material which is prescribed by the load applied at the end of the tube and is therefore treated as constant. In FPL we assume that $\zinf =1$, with an appropriate force to ensure this. For a tube with closed ends and no axial loading, we require that the force balance in the $Z$ direction is zero, and hence $C_2 =0$, giving the following relation from \rr{equib1},
\begin{equation}
\rinf \hat{W}_1(\rinf,\zinf) = 2 \zinf \hat{W}_2(\rinf,\zinf),\label{closedends}
\end{equation}
which may be used to determine $\zinf$ for any given $\rinf$. Therefore we take $\rinf$ as the controlling parameter of the deformation, with $P^{*}$ determined by \rr{pressure} and $\zinf$ either determined from \rr{closedends} or prescribed.

For examples and numerical results throughout this work we will use three strain-energy functions, the Varga, Ogden and Gent materials, given respectively by,
\begin{equation}
W=2
(\lambda_1+\lambda_2+\lambda_3 -3), \la{vargmat}
\end{equation}
\begin{equation}
W=\sum_{r=1}^{3} \mu_r
(\lambda_1^{\alpha_r}+\lambda_2^{\alpha_r}+\lambda_3^{\alpha_r}
-3)/\alpha_r, \la{ogdenmat}
\end{equation}
\begin{equation}
W=-\frac{1}{2} J_m \ln (1-\frac{\lambda_1^2 + \lambda_2^2 +\lambda_3^2 - 3}{J_m}), \la{gentmat}
\end{equation}
where we have nondimensionalised with respect to the infinitesimal shear modulus, $J_m>0$
is a material constant representing the maximum sustainable stretch of the material and $ \alpha_1=1.3, \alpha_2=5.0, \alpha_3=-2.0, \mu_1=1.491,  \mu_2=0.003, \mu_3=-0.023$. The Ogden and Gent materials were proposed in Ogden (1972) and Gent (1996) respectively, and are popularly used to model rubber. We include these three strain-energy functions as examples due to their popularity in the literature, though any suitable strain-energy function may be used.

The closed ends relation \rr{closedends} for the Varga and Gent materials become respectively,
\begin{equation}
1+ \rinf^2 \zinf - 2 \rinf \zinf^2 =0, \qquad 1 + \rinf^4 \zinf^2 - 2 \rinf^2 \zinf^4=0, \label{zinfexamples}
\end{equation}
which may be solved explicitly for $\zinf$. The counterpart of \rr{zinfexamples} for the Ogden material is more involved but it is found that all three materials display
a similar monotone relationship between $z_\infty$ and $r_\infty$  for $r_\infty>1$.  It is noted that the condition for the Gent material \rr{zinfexamples}$_2$ is independent of $J_m$. In fact, \rr{zinfexamples}$_2$ is valid for any strain energy that is only a function of the first invariant $I_1 = \lambda_1^2 + \lambda_2^2 + \lambda_3^2$, whereas
\rr{zinfexamples}$_1$ is valid for any strain energy that is only a function of $ \lambda_1  + \lambda_2  + \lambda_3$.

%
%

\section{Characterisation of solitary-wave type and kink-type solutions}\label{character}
\setcounter{equation}{0}

\begin{figure}[htf]
 \psfrag{aa}{$\rinf$}
 \psfrag{bb}{\hspace{-10pt}$r_0 -\rinf$}
\centering \epsfig{figure=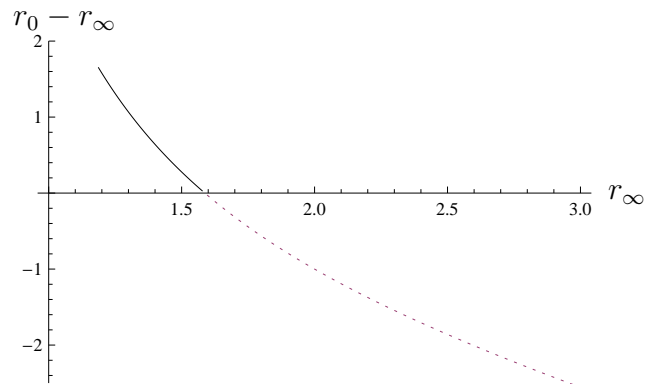,width=.5\textwidth}
\caption{Dependence of $r_0-\rinf$ on $r_\infty$ for the closed Varga tube. Only the solid line corresponds to localised solutions.}
\label{vargaclosed}
\end{figure}

\begin{figure}[htf]
 \psfrag{aa}{$\rinf$}
 \psfrag{bb}{\hspace{-10pt}$r_0 -\rinf$}
 \psfrag{cc}{$\rinf$}
 \psfrag{dd}{$r_0$}
 \psfrag{w}{(a)}
 \psfrag{x}{(b)}
\epsfig{figure=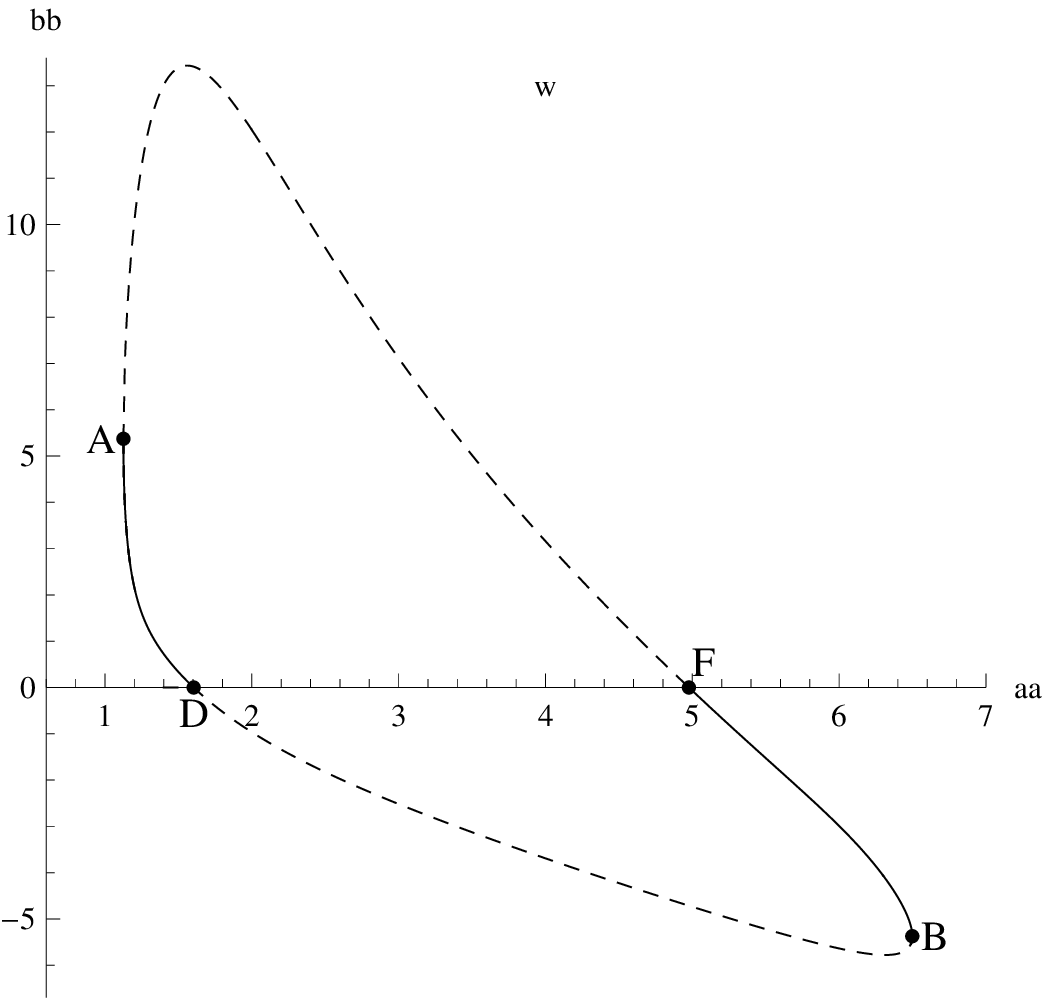,width=.45\textwidth}
\hspace{20pt}
\epsfig{figure=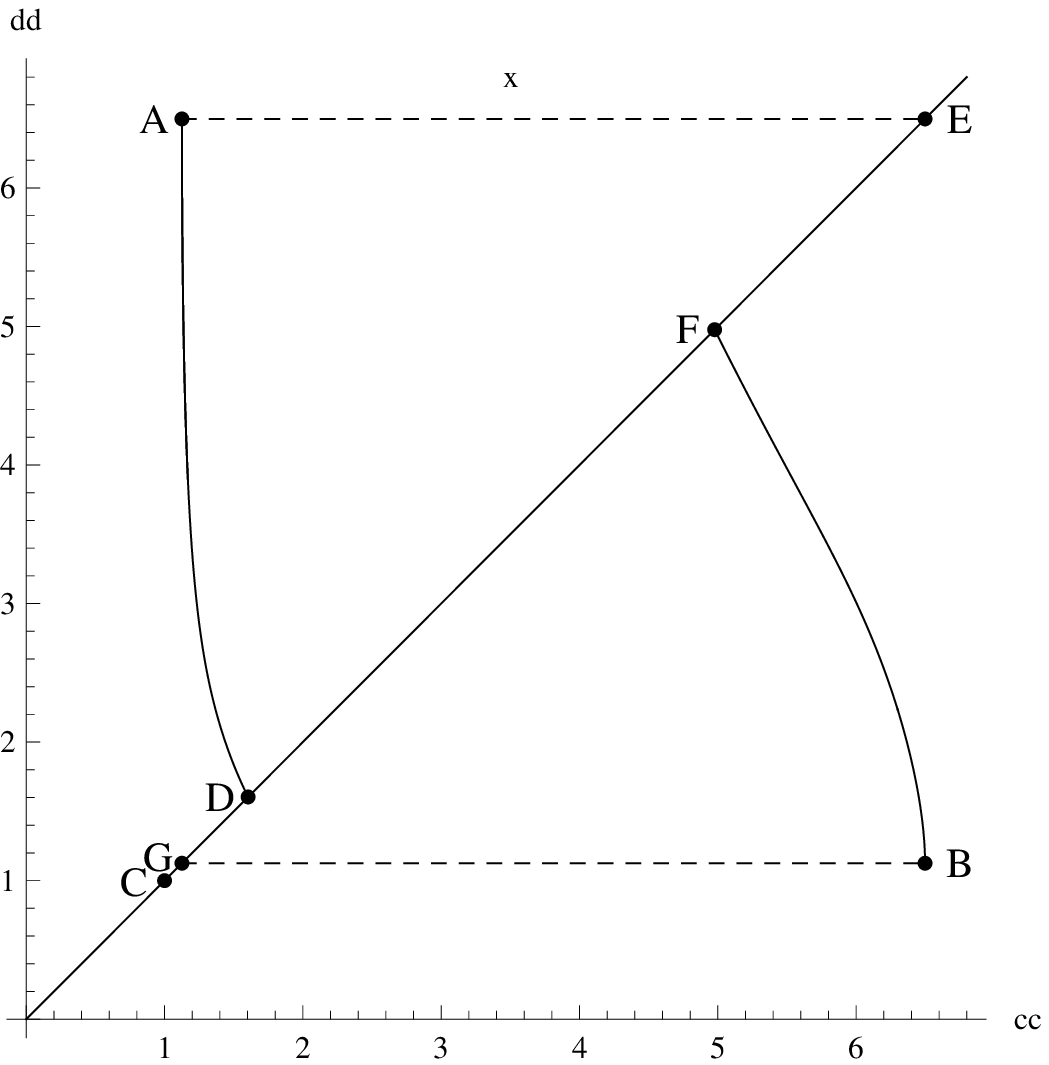,width=.45\textwidth}
\caption{Dependence of $r_0-\rinf$ and $r_0$ on $r_\infty$ for the closed Ogden tube. Only the solid lines in (a) correspond to localised solutions, and the segments corresponding to non-localised solutions are not plotted in (b)}
\label{ogdenclosed}
\end{figure}

\begin{figure}[htf]
 \psfrag{aa}{$\rinf$}
 \psfrag{bb}{\hspace{-10pt}$r_0 -\rinf$}
 \psfrag{cc}{$\rinf$}
 \psfrag{dd}{$r_0$}
  \psfrag{w}{(b)}
 \psfrag{x}{(a)}
\epsfig{figure=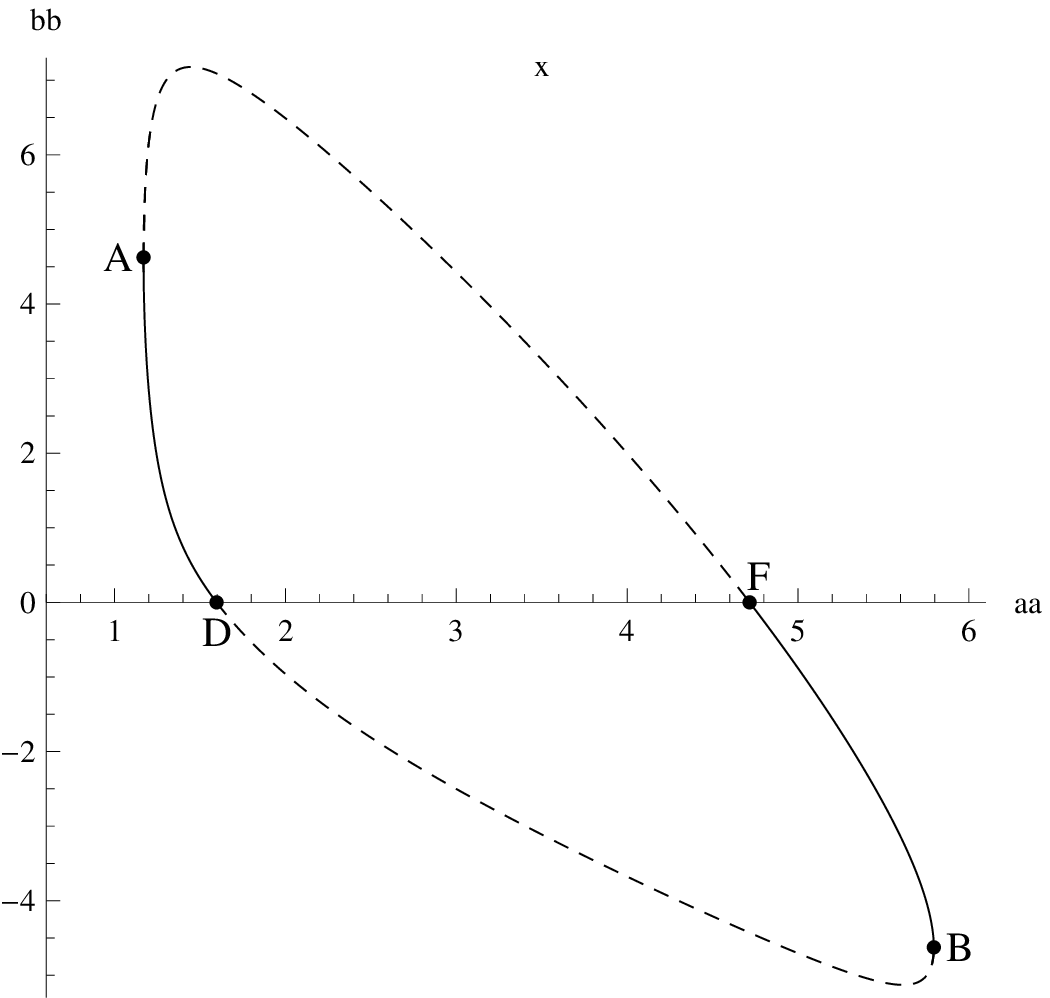,width=.45\textwidth}
\hspace{20pt}
\epsfig{figure=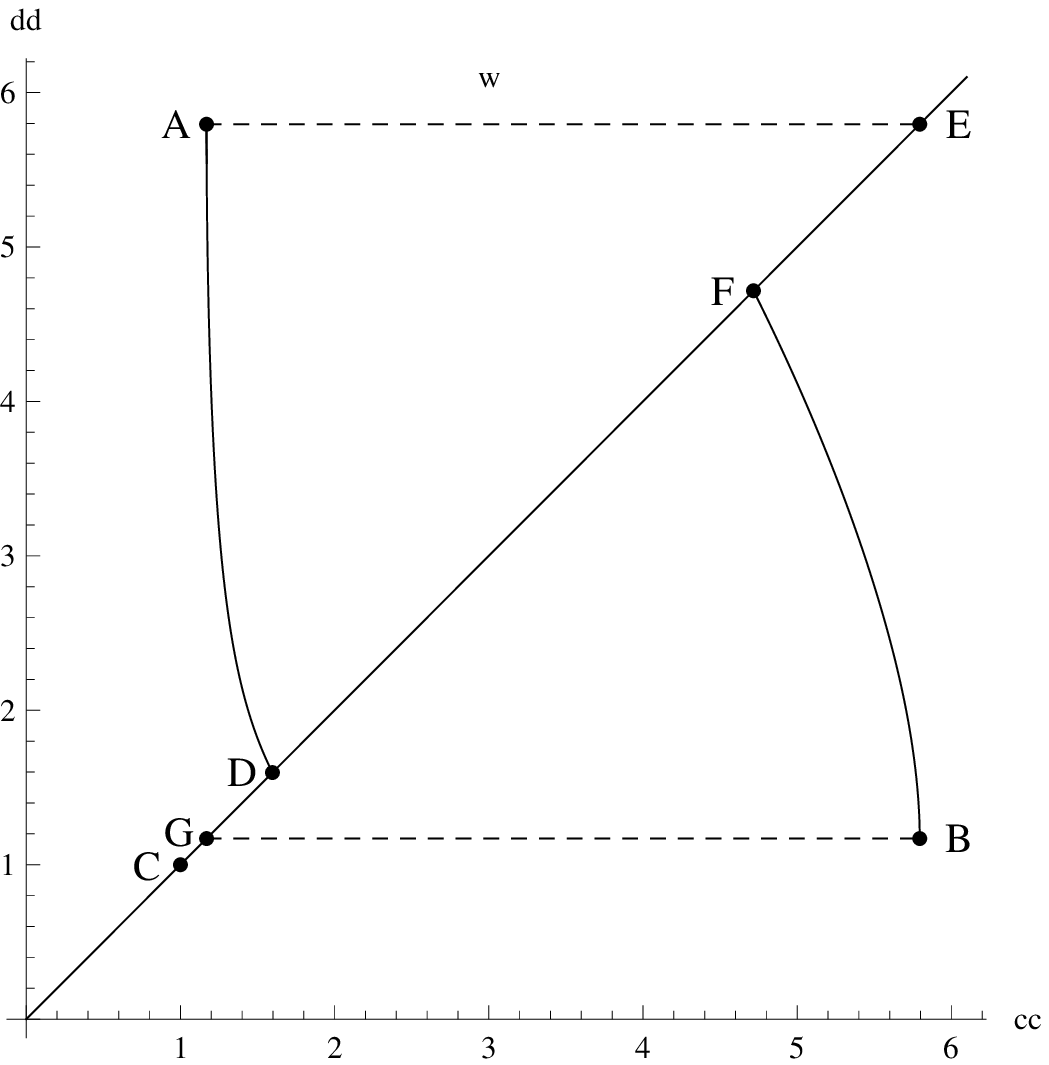,width=.45\textwidth}
\caption{Dependence of $r_0-\rinf$ and $r_0$ on $r_\infty$ for the closed Gent tube with $J_m =97.2$. Only the solid lines in (a) correspond to localised solutions, and
the segments corresponding to non-localised solutions are not plotted in (b)}
\label{gentclosed}
\end{figure}

Without loss of generality, we assume that the centre of the bulge/neck is located at $Z=0$, where we must necessarily have $r'(0)=0$ due to the
symmetry.
On evaluating \rr{equib2} and \rr{equib1} at $Z=0$, we obtain
\be \hat{W}(r_0, z'_{0})-z'_{0}\; \hat{W}_2(r_0,
z'_{0})-\hat{W}^{(\infty)}+ z_\infty \hat{W}_2^{(\infty)}=0, \la{r02} \en
\be
 \hat{W}_2(r_0, z'_{0})-\frac{\hat{W}_1^{(\infty)}}{2
r_\infty z_\infty} (r_0^2-r_\infty^2)-\hat{W}_2^{(\infty)}=0, \la{r01} \en
where $r_0=r(0), z'_{0} = z'(0) \ge 0$. Solving these two equations simultaneously for $ r_0$ and $z'_0$, we can obtain $r_0$ as a function of $\rinf$. As in FPL, we have
shown in Figures \ref{vargaclosed}, \ref{ogdenclosed}(a) and \ref{gentclosed}(a) $r_0-r_\infty$ versus $r_\infty$  for the Varga, Ogden and Gent strain-energy functions with closed ends. The corresponding plots for the case where $\zinf =1$ have previously been given in FPL, along with further discussion of the Varga material for this case. The analysis given in Section 5 of FPL for the Varga material still holds for the tube with closed ends, with minor adjustment of FPL's equation (5.7), in particular the fact that the equations blow up at a finite value of $\rinf =r^{*}= (4 (\sqrt{2} -1))^{1/3}$, at which point $z'_0 \to \infty$. For $\rinf<r^{*}$ there exist no non-trivial solutions with positive $z'_0$ for the closed tube.

We shall now focus our discussion on the Ogden and Gent strain-energy functions which are known to be realistic material models for rubber.
We first note that in each case the relation between $r_0-r_\infty$ and $\rinf$ is a closed curve that intersects the horizontal axis twice.
These two intersections are both bifurcation points.
Based on our numerical calculations, the near-critical analysis in FPL, and our further insight into the plots to be discussed shortly, we remark that only the solid line parts in these figures correspond to localised solutions. If we were to integrate the equations \rr{equib2} and \rr{equib1} from $Z=0$ using
values of $r_0$ and $z'_0$ from the other sections, we would obtain either unbounded or periodic solutions which do not satisfy our conditions at infinity. As shown in FPL, the smaller bifurcation value of $\rinf$
corresponds to a bifurcation into a bulging solution, whereas the larger bifurcation value corresponds to a bifurcation into a necking solution, which can readily be seen from the amplitude diagrams.

It will now be shown that the turning points $A$ and $B$ in Figures \ref{ogdenclosed}(a) and \ref{gentclosed}(a) have a special interpretation, namely that at these points we have
$r''(0)=0$ as well as $r'(0)=0$.
To this end, we first differentiate \rr{r02} and \rr{r01} with respect to $\rinf$, viewing $z'_{0}$ as a function of $r_0$
and $\rinf$, and $r_0, \zinf$ as functions of $\rinf$. By taking the limit $\paa r_0/\paa \rinf \to \infty$ in the resulting equations, we obtain
$$ \hat{W}_1(r_0, z'_{0})-z'_{0} \hat{W}_{12}(r_0, z'_{0})+z'_{0} \frac{\paa z'_{0}}{\paa r_0}\,\hat{W}_{22}(r_0, z'_{0})=0, $$ $$
\hat{W}_{12}(r_0, z'_{0}) -\frac{r_0 \hat{W}_{1}^{(\infty)}}{\rinf \zinf}+ \frac{\paa z'_{0}}{\paa r_0}\, \hat{W}_{22}(r_0, z'_{0})=0. $$
Finally, on eliminating $\paa z'_{0}/\paa r_0$ from the two equations above, we obtain
\be \hat{W}_1(r_0, z'_{0})-\frac{r_0 z'_{0}}{\rinf \zinf} \hat{W}_{1}^{(\infty)}=0. \la{ro3} \en
On the other hand, the static form of \rr{mot}$_2$, together with \rr{pressure}, may be rewritten as
\be
\left( \frac{\sigma_2}{\lambda_2^2}\right) r''+\left( \frac{\sigma_2}{\lambda_2^2}\right)' r'-\hat{W}_1+\frac{r_0 z'_{0}}{\rinf \zinf} \hat{W}_{1}^{(\infty)}=0. \la{ro4} \en
On evaluating this equation at $Z=0$ where $r'=0$, and making use of \rr{ro3}, we obtain $r''(0)=0$.

The result established above indicates that as we trace from the first bifurcation point along the solid curve the radius at the centre of
the bulge will increase monotonically until we reach the turning point $A$, where the bulge flattens out at its centre, stops
growing in radius and then starts to propagate in both directions; see Figure \ref{bulgedr}. At this stage the bulge can be viewed as two kink solutions stitched together
and each kink consists of two uniform states, $r=\rinf$ and $r=r_0$ respectively, joined by a smooth transition region. We now show that
these two uniform states in fact satisfy the so-called Maxwell equal-area rule (Ericksen 1975, Chater and Hutchinson 1984).

To show this, we first define a volume measure $v$,
\be v=\rinf^2 \zinf, \la{volume} \en
which for uniform inflation is the volume change per unit volume in the undeformed configuration. With the additional use of \rr{closedends}, we may
view $r_\infty$ and $z_\infty$ both as functions of $v$.

The Maxwell equal-area rule defines a pressure $P_m$ such that the two areas bounded by the curve $P(v)$ and the line $P=P_m$ are equal, i.e.
\begin{equation}
\int_{v_1}^{v_2} P(v) dv = P_m (v_2-v_1). \label{maxdef}
\end{equation}

\begin{figure}[!ht]
 \psfrag{aa}{$v$}
 \psfrag{bb}{$P^{*}$}
 \psfrag{u}{$v_1$}
 \psfrag{v}{$v_2$}
 \psfrag{w}{$\hspace{-10pt} P_m$}
\centering \epsfig{figure=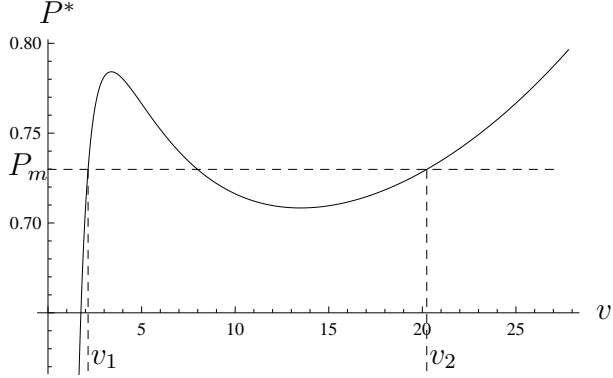,width=.5\textwidth}
\caption{Pressure as a function of volume for the closed Gent tube with $J_m=30$}
\label{presvol}
\end{figure}

The two values of $v$ thus generated are the volumes corresponding to the two uniform sections of the kinked solution. Figure \ref{presvol} shows the pressure-volume curve for a typical Gent tube with closed ends, along with the line $P_m$. This pressure-volume curve is typical for rubber-like materials, but the pressure-volume curve is not required to be non-monotonic for the kinked solution to exist, as discussed below.

We define the strain energy depending solely on the volume as $\tilde{W}(v)=\hat{W}(r_\infty(v), z_\infty(v))$.
It can then be shown that
\be P^*=2  \frac{d \tilde{W}}{d v}. \la{pv} \en
Thus, the Maxwell equal-area rule becomes
\be P_m (v_2-v_1)= 2 \left(\tilde{W}(v_2)-\tilde{W}(v_1)\right), \la{max} \en
where
\be v_1=\rinf^2 \zinf, \;\;\;\; v_2=r_0^2 z'_{0}, \;\;\;\; P_m=P^*|_{v=v_1}=P^*|_{v=v_2},\la{may} \en
with $(\rinf, \zinf)$ and $(r_0, z'_{0})$ being the two uniform states connected by the Maxwell line. It remains to show
that the $(r_0, z'_{0},\rinf,\zinf)$ defined in this way also satisfy the turning point condition \rr{ro3}.

For the case of closed ends we apply \rr{closedends} to both uniform states to obtain
\be \rinf \hat{W}_1(\rinf, \zinf) = 2 \zinf  \hat{W}_2(\rinf, \zinf), \;\;\;\;
r_0 \hat{W}_1(r_0, z'_{0}) = 2 z'_{0} \hat{W}_2(r_0, z'_{0}). \la{may3} \en
Also,  \rr{may}$_3$ may be written as
\be
P_m=\frac{\hat{W}_1(\rinf, \zinf)}{\rinf \zinf}. \la{may2} \en 
On substituting \rr{may}$_{1, 2}$ and \rr{may2} into \rr{max} and then making use of  \rr{equib2} and \rr{may3}, we do indeed obtain the turning point condition \rr{ro3}. It should be stressed that the equation \rr{may3} is only valid when considering a uniform state. For the case of fixed $\zinf$ \rr{may3} is not required but the turning point condition may still be derived.

For the case of the Gent tube with $\zinf=1$, $P$ is a monotonic function of $v$ and thus no Maxwell line exists. However, the condition given by \rr{ro3} still has a solution, corresponding to the kinked solution, as shown in Figure 3(a) of FPL.In this case, the two pressures given by \rr{may}$_3$ evaluated at the two pairs $(\rinf,1)$ and $(r_0,z'_0)$ are equal, and thus the Maxwell condition \rr{maxdef} is trivially satisfied.

A similar interpretation may be given to the  second turning point B, though this represents a kinked solution arising from a necking solution. To provide
further insight into the necking solution, we have shown in Figures \ref{ogdenclosed}(b) and \ref{gentclosed}(b) the corresponding $r_0$ against $r_\infty$, omitting
those segments that do not give rise to localized solutions. Viewed in this way,
point $B$ is simply a mirror reflection of $A$ about $r_0=r_\infty$, the line of uniform inflation. Thus, we may describe the entire inflation and/or deflation process as follows. First, the stress-free state corresponds to point $C$ in Figures \ref{ogdenclosed}(b) and \ref{gentclosed}(b). Uniform inflation would follow the straight line $r_0=r_\infty$ and terminate at the first
bifurcation point $D$. As inflation continues, the uniform configuration bifurcates into a bulged configuration; the growth of the bulge is described by the path $DA$. At point $A$, the bulge reaches its maximum and starts to propagate in both directions. For a finite tube, a uniform state will eventually be achieved as each of the two kinks reaches the end of the tube. This
uniform state corresponds to point $E$. At this stage, we may either inflate the tube further until it pops or deflate it. The deflation would follow the line $r_0=r_\infty$ until we reach
the second bifurcation point $F$. As deflation continues, the uniform state bifurcates into a necked state, the evolution of which is described by the path $FB$. At point $B$, the decrease
of the radius at the centre of the neck stops and the neck starts to propagate in both directions. The propagation stops when the kinks reach both ends, and the resulting new uniform state corresponds to point $G$. In the above description, we have assumed that the bulge or neck initiates in the middle of the tube. In practice, it is usually the material or geometrical inhomogeneity that selects the actual site of initiation.

Thus, the plots in Figures \ref{ogdenclosed} and \ref{gentclosed} are able to describe the entire bulging/necking process graphically. Figure \ref{gentclosed} is for the closed-end Gent tube with $J_m = 97.2$, but similar behaviour is found for $J_m > 18.23$, below which no bifurcation points exist. This is a larger value of $J_m$ than that found in FPL for the case of $\zinf=1$
(according to Horgan and Saccomandi 2003, the values of $J_m$ for healthy arteries range between 0.422 and 3.93). The Ogden tube with $\zinf =1$, discussed in FPL, has only one critical point and no kinked solution in contrast to the case of closed ends.

\section{Determination of the bulging/necking solutions}
\setcounter{equation}{0}
In this section, we use the phase plane method to provide a different perspective on how a bulging/necking solution evolves into a kink solution, and explain
how such solutions can be determined numerically. To this end, we
rewrite \rr{equib2} and \rr{equib1}, defining two new functions $f$ and $g$,
\begin{equation}
f(r,\lambda_{2}) \equiv \hat{W} -
\lambda_{2} \hat{W}_{2} -C_{1}=0, \label{f}
\end{equation}
\begin{equation}
g(r,\lambda_{2}) \equiv \frac{\lambda_{2}}{\hat{W}_{2}} (C_{2} +
\frac{P^{*}}{2}r^{2}) = z'.\label{g}
\end{equation}
Equation \rr{f} allows us to express $\lambda_2 = K(r)$ as a function of
$r$ for a given $\rinf$, though this relation will be
implicit for most strain-energy functions. Using the definition of $\lambda_2$ given in \rr{lambdas}, we can write
\begin{eqnarray}
(r')^2 &=& \lambda_{2}^2 - z'^2   \nonumber \\
&=& K(r)^2 - g(r, K(r))  \nonumber \\
&=& F(r;\rinf), \label{rdsqdeqn}
\end{eqnarray}
defining the function $F$. The behaviour of $F$ governs the existence and shape of the non-trivial solution. From elementary dynamical systems theory we may deduce that
a bulged solution can exist if $F$ has a double root at $r=r_\infty$, another
root at $r=r_0$ where $r_0>r_\infty$, and $F> 0$ for $r \in
(r_\infty, r_0)$. A similar statement with $r_0 <r_\infty$ can be made about necking solutions.

Following FPL, we may expand \rr{rdsqdeqn}  for values of $r$ close to $\rinf$ as
\begin{equation}
(r')^2= w'^{2} = \omega(\rinf) w^2 + \gamma(\rinf) w^3 + \mathcal{O}(w^4),\la{1.0}
\end{equation}
where $w = r- \rinf$, and the function $\omega$ is given by
\begin{equation}
\omega(\rinf) = \frac{\rinf
(\hat{W}_1^{(\infty)}-\zinf \hat{W}_{12}^{(\infty)})^2+\zinf^2 \hat{W}_{22}^{(\infty)}
(\hat{W}_1^{(\infty)}-\rinf \hat{W}_{11}^{(\infty)})}{\rinf \zinf \hat{W}_2^{(\infty)} \hat{W}_{22}^{(\infty)}}. \label{bifnew}
\end{equation}
The expression for $\gamma(\rinf)$ is too long and so is not written out here for brevity. As observed in FPL, the bifurcation condition
is given by $\omega(\rinf)=0$.

Equation \rr{1.0} confirms that in the near-critical regime where $|w|\ll 1$, the function $F$ always has a repeated root $r_\infty$ and one other root
approximately equal to $r_\infty -\omega(\rinf)/\gamma(\rinf)$.

On differentiating \rr{1.0} with respect to $Z$ we find,
\begin{equation}
w''=\omega(\rinf) w +\frac{3}{2} \gamma(\rinf) w^2+\mathcal{O}(w^3). \la{1.1}
\end{equation}
Expanding the above equation around $r_{cr}$, a root of the bifurcation condition $\omega(\rinf)=0$, defining $ \ep =\rinf-r_{cr}$, and then neglecting terms
of order higher than $\ep^2$, we obtain
\begin{equation}
w''= \omega'(r_{cr}) \epsilon w + \frac{3}{2} \gamma(r_{cr}) w^2,
\end{equation}or equivalently,
\begin{equation}
\frac{d^2V}{d \xi^2 }=V-   V^2, \la{1.4}
\end{equation}
where
\begin{equation}
  w=-\frac{2\ep \omega'(r_{cr})}{3 \gamma(r_{cr})} V(\xi), \;\;\;\; \xi=\sqrt{\ep \,\omega'(r_{cr})} Z. \la{fard}
\end{equation}
In writing down the last expression we have assumed that $\epsilon \omega'(r_{cr})>0$,
 which is a necessary condition for the existence of localised bulging or necking solutions as shown in FPL.
 Equation \rr{1.4} has an exact solitary-type solution given
 by
 \begin{equation}
 V= {V}_0 \equiv \frac{3}{2} {\rm sech}^2 (\frac{\xi}{2}), \la{soli}
 \end{equation}
 which will be referred to as the {\it weakly nonlinear solution}. We observe  from \rr{fard} and the definition $w = r- \rinf$ that this solution corresponds to
 a localized bulging solution
 if  $\gamma(r_{cr})<0$, and to a localized
 necking solution  if $\gamma(r_{cr})> 0$.

\begin{figure}[!ht]
 \psfrag{"}{}
 \psfrag{v}{$r$}
 \psfrag{w}{(a)}
 \psfrag{x}{(b)}
 \psfrag{y}{(c)}
 \psfrag{z}{(d)}
 \psfrag{ii}{$F(r;1.5)$}
 \psfrag{jj}{$F(r;1.11694)$}
 \psfrag{kk}{$r'$}
 \psfrag{ll}{$r'$}
\centering \epsfig{figure=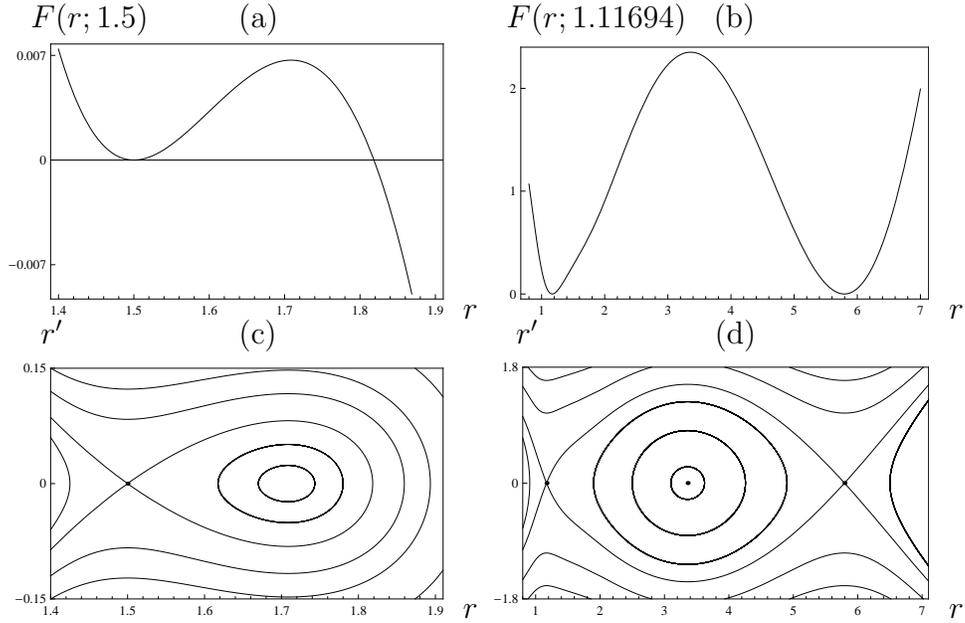,width=.8\textwidth}
\caption{Plots of $F$ against $r$ for the closed Gent tube with $J_m=97.2$ for $\rinf$ corresponding to (a) a typical bulged solution and (b) the kinked solution, above the corresponding phase portraits.}
\label{allfour}
\end{figure}

Return now to the fully nonlinear equation \rr{rdsqdeqn} which, when differentiated with respect to $Z$, yields $2 r''= \paa F/\paa r$. Thus, fixed points are given by the roots of
$\paa F/\paa r=0$ and on the phase plane there exist saddles at the minima of $F$ and centres at the maxima. For $r_m <r_\infty <r_{cr}$, where $r_{cr}$ is the first bifurcation value and $r_m$ is
the value of $r_\infty$ corresponding to the turning point $A$ in Figures \ref{ogdenclosed}(a) and \ref{gentclosed}(a), the profile of $F(r;\rinf)$ and the corresponding phase portrait
are typified by those shown in Figures \ref{allfour}(a, c). In this case, $F$ has a repeated root $r_\infty$ and a third root $r_0$, and we have a localized bulging solution corresponding to the homoclinic orbit in \ref{allfour}(c). In the limit $r_\infty \to r_m$,
the third root $r_0$, by coalescing with a fourth root, becomes another double root and a local point of minimum of $F$; see Figure \ref{allfour}(b). In this case we have a kink solution corresponding to the heteroclinic orbit in Figure \ref{allfour}(d).

 %

\begin{figure}[!ht]
\psfrag{aa}{Z}
\psfrag{bb}{r(Z)}
\centering \epsfig{figure=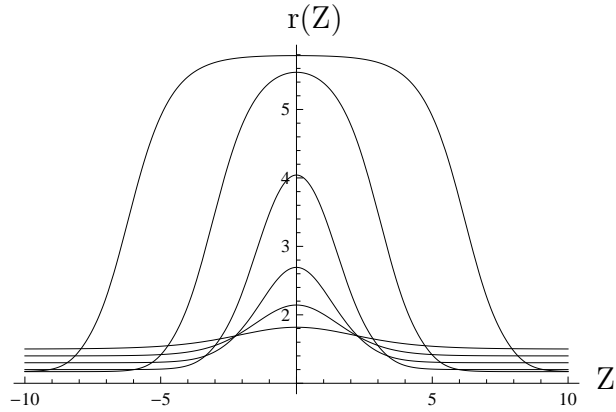,width=.5\textwidth}
\caption{Profiles of the bulge corresponding to $\rinf = 1.16941,1.17,1.2,1.3,1.4,1.5$ for the closed Gent tube with $J_m =97.2$. Larger amplitudes of $r(Z)$
correspond to smaller values of $\rinf$.}
\label{bulgedr}
\end{figure}

There are two methods that may be used to find the {\it fully nonlinear} solitary-wave type solution corresponding to the homoclinic orbit. The first method, as discussed in Section 5 of FPL, involves numerical integration of a system of three first-order differential equations. The second method is rewriting \rr{rdsqdeqn} as,
\begin{equation}
\int_{r_0}^{r(Z)} \frac{dr}{-\sqrt{F(r;\rinf)}} = \int_{0}^{Z} dZ=Z, \;\;\;\; Z>0,  \label{intrs}
\end{equation}
where we have used the fact that $r'(Z)<0$ for $Z>0$. When evaluated numerically, this equation gives $Z$ as a function of $r$. When a symbolic algebra package such as {\it Mathematica} is used,
the inversion to find $r$ as a function of $Z$ can be carried out simply by exchanging two columns of data. We have used both methods to validate our numerical results. Figure \ref{bulgedr} shows typical profiles of the solitary-wave type solution for different $\rinf$ for the closed Gent tube. In particular it shows how the solution stops growing radially and begins to propagate down the tube as the turning point $A$ in Figure \ref{gentclosed}(a) is approached.

\section{Stability of the weakly nonlinear solution}
\setcounter{equation}{0}
We now consider stability of the weakly nonlinear solitary-wave type solution given by \rr{soli}. The result will be used to validate our stability analysis of the fully nonlinear
solution to be presented in the next section.

The static solution \rr{soli} is in fact a \lq fixed'  point of an evolution equation when time dependence is included.
For convenience we now consider the necking case where $\ep$ and $\omega'(r_{cr})$ are both positive; exactly the same analysis applies for the bulging case where $\ep$ and $\omega'(r_{cr})$ are both negative though absolute values then need to be taken throughout the following section.

For $\ep \ll 1$, the prestressed membrane tube will support traveling waves with small wave number and small wave speed. It can be shown with the aid of the dispersion relation derived in Fu and Il'ichev (2009) that the wave number and wave speed are of  order $\sqrt{\ep}$ and $\ep$, respectively. It can also be deduced that
the radial amplitude is of the order $\sqrt{\ep}$ times the axial amplitude.
Thus, we may define
a far distance variable $\xi$ as in \rr{fard}, a slow time variable $\tau$ through
\be \tau=\ep t, \la{slowt} \en
and look for a perturbation solution of the form
\be
\rinf=r_{cr}+\ep \hat{\lambda}_1, \;\;\;\; \zinf=z_{cr}'+\ep \hat{\lambda}_2, \la{prii}
\en
\begin{equation}
r=\rinf +\ep \left\{ w_1(\xi, \tau)+\ep\, w_2(\xi, \tau)+ \cdots \right\},
\end{equation}
\begin{equation}
z=\zinf Z +\sqrt{\ep} \left\{ u_1(\xi, \tau)+\ep\, u_2(\xi, \tau)+ \cdots \right\}, \la{pert}
\end{equation}
where  $\hat{\lambda}_1$ and $\hat{\lambda}_2$ are constants, and $w_1, w_2, u_1, u_2$ etc are to be determined.

The internal pressure is given by
$$ P^{*}=\frac{W_1(\rinf, \zinf)}{\rinf \zinf}, $$
and we assume that it is held fixed in any axisymmetric perturbations. This is known as pressure controlled inflation
which can be realised by connecting the gas in the tube to a very large reservoir of the same gas. We note, however, that
with $\rinf$ and $\zinf$ given by \rr{prii}, we have the Taylor expansion
\be P^{*}=P_0+\ep P_1+\cdots. \la{eqnp} \en
On substituting \rr{prii}--\rr{eqnp} into the equations of motion \rr{mot} and equating the coefficients of like powers of $\ep$, we obtain,
to leading order,
\be
L \left[\begin{array}{c} w_1 \\ \sqrt{\omega'(r_{cr})} \,u_{1\xi} \end{array}\right]=0, \;\;\;\;
L= \left[\begin{array}{cc} -\hat{W}_1/z_{cr}'+\hat{W}_{12} & \hat{W}_{22}  \\ z_{cr}' (\hat{W}_1-r_{cr} \hat{W}_{11}) &
r_{cr} (\hat{W}_1-z_{cr}' \hat{W}_{12})
 \end{array}\right], \la{order1} \en
 where $\hat{W}_1, \hat{W}_2, \hat{W}_{12}, \hat{W}_{22}$ are all evaluated at $r=r_{cr}, \; z'=z_{cr}'$, and $u_{1\xi}$ denotes
 $\paa u_1/\paa \xi$. It is easy to find $ \det L = \omega(r_{cr}) r_{cr} \hat{W}_2 \hat{W}_{22}$. Thus, as we expected, $\omega(r_{cr})=0$
ensures that the matrix equation \rr{order1}$_1$ has a non-trivial solution for $w_1$ and $u_1$.

Proceeding to the next order, we find that $w_2$ and $u_2$ satisfy the inhomogeneous system
\be
L \left[\begin{array}{c} w_2 \\ \sqrt{\omega'(r_{cr})} \,u_{2\xi} \end{array}\right]={\bm b}, \la{order2} \en
where the vector ${\bm b}$ only contains $w_0$ and its derivatives. Forming the dot product of \rr{order2} with the left
eigenvector of $L$, we then obtain the evolution equation in the form
\be
\frac{\paa^2 V}{\paa \xi^2}-c_1 \frac{\paa^2 V}{\paa \tau^2}=c_2 \frac{\paa^4 V}{\paa \xi^4}+c_3 \frac{\paa^2 V^2}{\paa \xi^2}, \la{evo0} \en
where $c_1, c_2, c_3$ are known constants, and $V$ is given by
$$w_1=-\frac{2 \omega'(r_{cr})}{3 \gamma(r_{cr})} V(\xi, \tau) $$ which may be compared with its static form \rr{fard}$_1$.
Although the expressions for the constants $c_1, c_2, c_3$ are available from
the above perturbation procedure, we may obtain their expressions more simply as follows.

First, from the fact that when $V$ is assumed to be independent of $\tau$, \rr{evo0} must reduce to the static amplitude equation
\rr{1.4}, we deduce that $c_2=c_3=1$. To determine the remaining constant $c_1$,
we linearise \rr{evo0} and then look for a traveling wave solution of the form
\be V={\rm e}^{\ii K (\xi- v \tau)}={\rm exp} \left( \ii K \sqrt{\ep \,\omega'(r_{cr})} \left(Z-\sqrt{\frac{\ep}{\omega'(r_{cr})}} v t \right) \right),\la{1.6} \en obtaining,
\be v^2=\frac{1+K^2}{c_1 }.\en
From \rr{1.6}$_2$ we see that the actual wave number $\hat{k}$ and speed $\hat{c}$, using the notation of Fu and Il'ichev (2009), are
\be \hat{k}=K \sqrt{\ep \,\omega'(r_{cr})}, \;\;\;\; \hat{c}=v \sqrt{\frac{\ep}{\omega'(r_{cr})}}. \la{1.7a} \en
It then follows that
\be \hat{c}^2= \frac{\ep}{c_1 \omega'(r_{cr})}+\frac{\hat{k}^2}{c_1 \omega'(r_{cr})^2}
=\frac{(\rinf-r_{cr})}{c_1 \omega'(r_{cr})}+\frac{\hat{k}^2}{c_1 \omega'(r_{cr})^2}
. \la{1.7} \en
From equation (2.11) of Fu and Il'ichev (2009) we obtain
\be \frac{\rho \hat{c}^2}{\mu}=f(\rinf)+O(\hat{k}^2)=f'(r_{cr})(\rinf-r_{cr})+O(\hat{k}^2, \ep^2), \la{1.8} \en
 where
$$ f(\rinf)= - \frac{\hat{W}_2^{(\infty)} \hat{W}_{22}^{(\infty)}}{\zinf \left(\hat{W}_{1}^{(\infty)}-\rinf  \hat{W}_{11}^{(\infty)}\right)} \omega(\rinf) $$
The $c_1$ is then obtained by comparing \rr{1.7} with \rr{1.8}. We have
\begin{equation}
\frac{c_1}{\rho}=\frac{\zinf}{\rinf} \cdot \frac{\rinf \hat{W}_{11}^{(\infty)}- \hat{W}_{1}^{(\infty)}}{\hat{W}_2^{(\infty)}\hat{W}_{22}^{(\infty)}}
\cdot \frac{1}{\omega'(r_{cr})^2}, \la{c1}
\end{equation}
where the right hand side is evaluated at the bifurcation point. Therefore our evolution equation is given by,
\begin{equation}
\frac{\paa^2 V}{\paa \xi^2}-c_1 \frac{\paa^2 V}{\paa \tau^2}=\frac{\paa^4 V}{\paa \xi^4}+\frac{\paa^2 V^2}{\paa \xi^2}, \la{evo2}
\end{equation}
with $c_1$ given by \rr{c1}. We note from \rr{c1} that $c_1$ is non-negative for values of $(r_\infty, r_0)$ on the solid segments in Figures 2-4.
Equation \rr{evo2} is recognized as a Boussinesq equation whose solution has been much studied; see, e.g., Ablowitz and Clarkson (1991).

To study the stability of \rr{soli}, we substitute
$$ V=V_0(\xi)+B(\xi) {\rm e}^{\sigma \tau} $$
into \rr{evo2} and linearise to obtain
\begin{equation}
\frac{d^4 B}{d \xi ^4}- \frac{d^2 B}{d \xi ^2}+2 \frac{d^2 (V_0 B)}{d \xi ^2}+c_1 \sigma^2 B=0. \la{eig0}
\end{equation}
The static solution $V_0(\xi)$ is said to be linearly unstable or spectrally unstable if, for some fixed complex $\sigma$ with ${\rm Re}\,(\sigma) >0$, there exists a solution of
\rr{eig0} which decays exponentially as $\xi \to \pm \infty$.

The above eigenvalue problem is now solved by computing the Evans function. The Evans function is a complex analytic function whose zeros correspond
to the eigenvalues (Evans 1975; Alexander {\it et al}. 1990). We follow the procedure explained in Afendikov and Bridges (2001) in which the eigenvalue
problem also involves a fourth-order differential equation. We note that the Boussinesq equation \rr{evo2} also admits a solitary wave solution of the
form $V_c(\xi-c  \tau)$, which reduces to $V_0$ when $c=0$. The stability of this solitary wave solution has previously been studied by
Alexander and Sachs (1995), also with the Evans function method. They normalized their Evans function $E(\sigma)$ such that it tended to unity as $\sigma \to \infty$.
They further showed that $E(0)=E'(0)=0$, and then with the use of an explicit expression for $E''(0)$, they deduced that the solution $V_c(\xi-c  \tau)$ was unstable if
$c < 1/(2 \sqrt{c_1})$. It then follows immediately that our static solution $V_0$ is unstable. Despite this known result, we shall still use this
simple case to illustrate how the Evans function
can be calculated. Our procedure is different from that of Alexander and Sachs (1995) and will be used in the determination of stability of the fully nonlinear solution in the next
section. Furthermore, an exact solution that will emerge from such a calculation seems be new.

We rewrite the system \rr{eig0} as a system of first order differential equations
\begin{equation}
\bm{y}' = A(\xi; \zeta) \bm{y}, \label{compound1}
\end{equation} where
\begin{equation}
\bm{y}= \left(\begin{array}{c}B(\xi) \\ B'(\xi) \\B''(\xi) \\B'''(\xi)
            \end{array}
          \right), \;\;\;\;     A(\xi; \zeta) = \left(\begin{array}{cccc}
              0 & 1 & 0 & 0 \\
              0 & 0 & 1 & 0 \\
              0 & 0 & 0 & 1 \\
              - \zeta -2 V_0''(\xi) & -4 V_0'(\xi) & 1-2 V_0(\xi) & 0 \\
            \end{array}
          \right),
\end{equation}
 and $\zeta = c_1 \sigma^2$. As $\xi \to \pm \infty$, $A$ has two pairs of eigenvalues, given by $\pm k_1, \pm k_2$, where $$k_1 = \sqrt{\frac{1}{2} (1-\sqrt{1-4 \zeta})}, \;\;\;\;
k_2 = \sqrt{\frac{1}{2} (1+\sqrt{1-4 \zeta})},$$
with positive square root taken on all four occasions.

We denote the eigenvectors associated with $-k_1, -k_2, k_1, k_2$ by ${\bm a}^+_1, {\bm a}^+_2, {\bm a}^-_1, {\bm a}^-_2,$ respectively. Then as $\xi \to \infty$, any decaying
solution of \rr{compound1} will tend to a linear combination of ${\bm a}^+_1 {\rm e}^{-k_1 \xi}$ and ${\bm a}^+_2 {\rm e}^{-k_2 \xi}$. Likewise, as $\xi \to -\infty$, any decaying
solution of \rr{compound1} will tend to a linear combination of ${\bm a}^-_1 {\rm e}^{k_1 \xi}$ and ${\bm a}^-_2 {\rm e}^{k_2 \xi}$. Choosing $l$ to be a suitably large positive number,
we may integrate \rr{compound1} subjected to the initial conditions
$$ {\bm y}(l)={\bm a}^+_1, \;\;\;\; {\bm y}(l)={\bm a}^+_2, \;\;\;\;{\bm y}(-l)={\bm a}^-_1, \;\;\;\;{\bm y}(-l)={\bm a}^-_2, $$
in turn to obtain four independent solutions
$$ {\bm y}^+_1(\xi), \;\;\;\;{\bm y}^+_2(\xi),\;\;\;\; {\bm y}^-_1(\xi), \;\;\;\;{\bm y}^-_2(\xi).$$
It then follows that any solution of \rr{compound1} that decays as $\xi \to \infty$ must take the form $d_1 {\bm y}^+_1(\xi)+d_2 {\bm y}^+_2(\xi)$, where $d_1, d_2$ are
constants. Likewise, any solution of \rr{compound1} that decays as $\xi \to- \infty$ must take the form $d_3 {\bm y}^-_1(\xi)+d_4 {\bm y}^-_2(\xi)$. At an eigenvalue of $\zeta$, these two
solutions intersect at any specific $\xi$, say $\xi=d$. Thus, the eigenvalues may be determined from the condition $N(\zeta, d)=0$, where
\be N(\zeta, d)= \det [ {\bm y}^-_1(d),\;{\bm y}^-_2(d), \;{\bm y}^+_1(d), \;{\bm y}^+_2(d)]. \la{detn} \en
The above determinant is, in general, dependent on the matching point $d$, although the eigenvalues should be independent of it. The Evans function, $D(\zeta)$, is defined
by
\begin{equation}
D(\zeta)=e^{-\int^{d}_{-\infty} {\rm Tr} A\, d \xi} N(\zeta, d),
\end{equation}
and is independent of the matching point $d$ (this can be established with the use of the well-known formula
 $d (\det M)/d x = (\det M) \textrm{tr} \left( M^{-1}  d M/d x \right)$ for any square matrix function $M(x)$).

The above procedure breaks down when $k_1 = k_2$, that is when $\zeta \to 1/4$. In this case, we need to replace ${\bm a}^+_2, {\bm a}^-_2$ by the corresponding generalised
eigenvectors
\be \lim_{k_2 \to k_1} \frac{{\bm a}^+_2-{\bm a}^+_1}{k_2 - k_1}, \;\;{\rm and}\;\; \lim_{k_2 \to k_1} \frac{{\bm a}^-_2-{\bm a}^-_1}{k_2 - k_1}, \la{gene} \en
respectively. To accommodate this isolated case, we may replace the determinant in \rr{detn} by
$$ \det [ {\bm y}^-_1(d), \;\frac{{\bm y}^-_2(d)-{\bm y}^-_1(d)}{k_2-k_1},\; {\bm y}^+_1(d),\; \frac{{\bm y}^+_2(d)-{\bm y}^+_1(d)}{k_2-k_1}], $$
which is simply $1/(k_2-k_1)^2$ times the original determinant. Thus, equivalently, to take care of the above special case, we only need to use $N(\zeta, d)$ in the form
\be N(\zeta, d)= \frac{4}{4 \zeta-1} \det [ {\bm y}^-_1(d),\; {\bm y}^-_2(d), \;{\bm y}^+_1(d), \;{\bm y}^+_2(d)], \la{detn1} \en
which has a finite limit when $\zeta \to 1/4$.

To avoid any \lq\lq stiff" behaviour,
we shall use the compound matrix method (Gilbert and Backus 1966; Ng and Reid 1979, 1985) to evaluate the determinant $N(\zeta, d)$ and hence the Evans function.
To this end, we introduce two new matrices $\bm{Y}^{+}(\xi)$ and $\bm{Y}^{-}(\xi)$ through
 $$ \bm{Y}^{+}(\xi)=[{\bm y}^+_1(\xi), {\bm y}^+_2(\xi)], \;\;\;\; \bm{Y}^{-}(\xi)=[{\bm y}^-_1(\xi), {\bm y}^-_2(\xi)]. $$
 We then define the minors of these matrices, $\phi_i^{-}$ and $\phi_i^{+}, i = 1,\ldots 6$, as the determinants formed by taking the $(1,2),(1,3),(1,4),(2,3),(2,3),(3,4)$-th rows.
The two vector functions formed from these minors satisfy the differential equations
\begin{equation}
\frac{d \bm{\phi}^+}{d \xi} = Q(\xi) \bm{\phi}^+, \;\;\;\;\frac{d \bm{\phi}^-}{d \xi} = Q(\xi) \bm{\phi}^-, \label{Qphi}
\end{equation}
where
\begin{equation}
Q (\xi)= \left(
  \begin{array}{cccccc}
    A_{11}+A_{22} & A_{23} & A_{24} & -A_{13} & -A_{14} & 0 \\
    A_{32} & A_{11}+A_{33} & A_{34} & A_{12} & 0 & -A_{14} \\
    A_{42} & A_{43} & A_{11}+A_{44} & 0 & A_{12} & A_{13} \\
    -A_{31} & A_{21} & 0 & A_{22}+A_{33} & A_{34} & -A_{24} \\
    -A_{41} & 0 & A_{21} & A_{43} & A_{22}+A_{44} & A_{23} \\
    0 & -A_{41} & A_{31} & -A_{42} & A_{32} & A_{33}+A_{44} \\
  \end{array}
\right),
\end{equation}
see, for instance, Gilbert and Backus (1966) or Bridges (1999).
The initial conditions for $\bm{\phi}^+$ and $\bm{\phi}^-$ are given by the corresponding minors of $\bm{Y}^{+}(l)$ and $\bm{Y}^{-}(-l)$, respectively. Equation \rr{Qphi}$_1$ is then integrated from $\xi=l$ and \rr{Qphi}$_2$  from $\xi=-l$ for a given $\zeta$. However, to remove the exponential growth we write
$${\bm \phi}^+(\xi) = {\bm \psi}^+(\xi) e^{ -(k_1+k_2) \xi}, \;\;\;\;{\bm \phi}^-(\xi) = {\bm \psi}^-(\xi) e^{ (k_1+k_2) \xi}, $$ so that
\begin{equation}
\frac{ d \bm{\psi}^+}{d \xi} = (Q+ (k_1+k_2) I) \bm{\psi}^+,  \;\;\;\; \frac{ d \bm{\psi}^-}{d \xi} = (Q- (k_1+k_2) I) \bm{\psi}^-, \label{Qphi2}
\end{equation}
where $I$ is the $6 \times 6$ identity matrix.

\begin{figure}[!ht]
 \psfrag{aa}{$\zeta$}
 \psfrag{bb}{$D(\zeta)$}
\centering \epsfig{figure=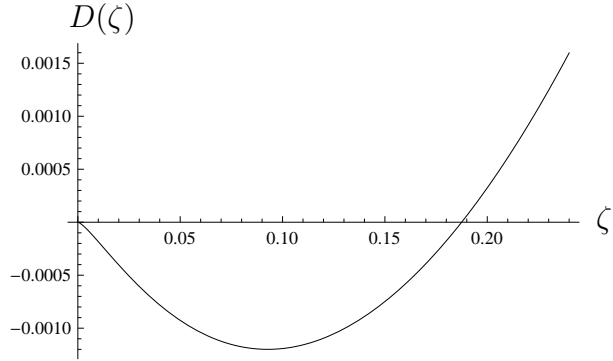,width=.5\textwidth}
\caption{The Evans function for the weakly nonlinear solution.}
\label{weakevans}
\end{figure}

\begin{figure}[!ht]
 \psfrag{aa}{$\xi$}
 \psfrag{bb}{$B(\xi)$}
\centering \epsfig{figure=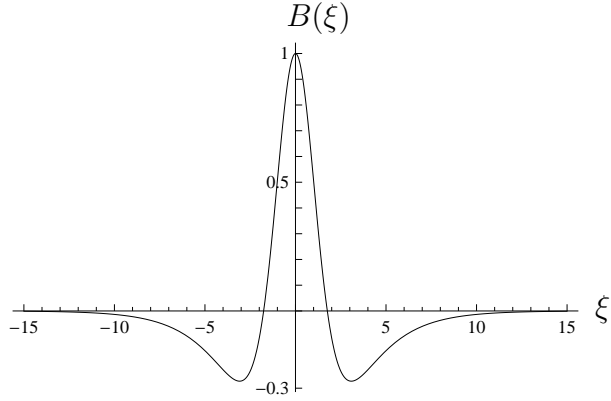,width=.5\textwidth}
\caption{Normalised eigenfunction of \rr{eig0} corresponding to the eigenvalue $\zeta_1 = 3/16$.}
\label{eigenfunction1}
\end{figure}

In terms of the vector functions $\bm{\psi}^+$ and $\bm{\psi}^-$, the determinant appearing in the definition of the Evans function becomes
$$
N(\zeta, d)=\frac{4}{4 \zeta-1} \left\{ \psi^{+}_1 (d) \psi^{-}_6 (d) -\psi^{+}_2 (d) \psi^{-}_5 (d) +\psi^{+}_3 (d) \psi^{-}_4 (d)\right.$$
\begin{equation}
\left.
+\psi^{+}_4 (d) \psi^{-}_3 (d) -\psi^{+}_5 (d) \psi^{-}_2 (d) +\psi^{+}_6 (d) \psi^{-}_1 (d) \right\}.
\end{equation}
We used the above procedure to calculate $D(\zeta)$ numerically. The results are shown in Figures \ref{weakevans} and \ref{eigenfunction1}. It is seen that $D(\zeta)=0$ has a single positive root, seemingly equal to the rational number $3/16$.   This suggests that the eigenvalue problem \rr{eig0} may have an exact solution. A systematic procedure for obtaining exact solutions is to
substitute a trial solution of the form
$$ B(\xi)=f(\xi) \left(d_1 {\rm sech} (\frac{\xi}{2})+ d_2 {\rm sech}^2 (\frac{\xi}{2})+d_3 {\rm sech}^3 (\frac{\xi}{2})
+d_4 {\rm sech}^4 (\frac{\xi}{2})\right) $$
into the differential equation \rr{eig0} and solving the resulting equations for the disposable constants $d_1, d_2, d_3$ and $d_4$.
For instance, for $f(\xi)=1,$ or ${\rm tanh}(\frac{\xi}{2})$, non-trivial solutions are always found despite the fact that the system of equations for $d_1, d_2, d_3$ and $d_4$ is over-determined. By taking $f(\xi)=1$, we find the exact
solution
$$ \zeta=\frac{3}{16}, \;\;\;\; B(\xi)=\textrm{sech}(\frac{\xi}{2}) -2 \, \textrm{sech}^{3}(\frac{\xi}{2}), $$
whereas by taking $f(\xi)={\rm tanh}(\frac{\xi}{2})$, we recover the exact solution $\zeta=0, \; B=V_0'(\xi)$, which could have
been deduced from the translational invariance of \rr{eig0}.

Therefore, as the above analysis is conducted for a general strain-energy function and valid for both bulging and necking solutions, we conclude that all near-critical
solitary-wave type solutions are unstable with respect to axisymmetric perturbations.

\section{Stability of the fully nonlinear solution}
\setcounter{equation}{0}
In this section we consider the stability of the fully nonlinear bifurcated solutions $r= \bar{r}(Z), \; z=\bar{z}(Z)$ that were determined in Section 4.
As in the previous section, we consider axisymmetric perturbations and write
\begin{equation}
r(Z, t) = \bar{r}(Z) + w(Z,t),\;\;\;\; z(Z, t) = \bar{z}(Z) + u(Z,t). \label{perturb1}
\end{equation}  On substituting \rr{perturb1} into \rr{mot} and linearising in terms of $w$ and $u$, we find
\begin{equation}
\left[\frac{\bar{\lambda}_{2}^2 \bar{W}_2 u' + \bar{z}' (\bar{\lambda}_{2} \bar{W}_{22} - \bar{W}_2)(\bar{r}' w' + \bar{z}' u') + \bar{\lambda}_{2}^2  \bar{z}' \bar{W}_{12} w}{\bar{\lambda}_2^3}\right]' \\
- P^{*} (\bar{r} w' + w \bar{r}') = \rho \ddot{u} \label{perturb2}
\end{equation}
\begin{equation}
\left[\frac{\bar{\lambda}_{2}^2 \bar{W}_2 w' + \bar{r}' (\bar{\lambda}_{2} \bar{W}_{22} - \bar{W}_2)(\bar{r}' w' + \bar{z}' u') + \bar{\lambda}_{2}^2 \bar{r}'  \bar{W}_{12} w}{\bar{\lambda}_{2}^3}\right]' \nonumber
\end{equation}
\begin{equation}
- \frac{\bar{W}_{12}}{\bar{\lambda}_{2}} (\bar{r}' w' + \bar{z}' u') - w \bar{W}_{11} - P^{*} (\bar{r} u' + w \bar{z}') = \rho \ddot{w}, \label{perturb2b}
\end{equation}
where $\bar{\lambda}_2=\sqrt{\bar{r}'^2+\bar{z}'^2}$, $\bar{W}_2=\hat{W}_2(\bar{r}, \bar{\lambda}_2)$, $\bar{W}_{12}=\hat{W}_{12}(\bar{r}, \bar{\lambda}_2)$
 etc.
It should be noted that $P^{*}$ in \rr{perturb2} and \rr{perturb2b} is a function of $\rinf$.

In the spectral stability analysis, we look for a solution of the form
\begin{equation}
 w(Z,t) = \tilde{w} (Z) e^{\eta t}, \qquad u(Z,t) = \tilde{u}(Z) e^{\eta t}. \label{perturb3}
\end{equation}
The fully nonlinear solution $r= \bar{r}(Z), \; z=\bar{z}(Z)$ is said to be linearly unstable or spectrally unstable if, for some fixed complex $\eta$ with ${\rm Re}\,(\eta) >0$, there exists a solution of the above form which decays exponentially as $Z \to \pm \infty$.

It can be seen that \rr{perturb2} and \rr{perturb2b}, after use of \rr{perturb3}, is a system of two coupled linear second order differential equations for $\tilde{w}(Z)$ and $\tilde{u}(Z)$, and the dependence on $\eta$ is entirely through the combination
\be \alpha =\rho \eta^2, \la{alpha} \en which defines $\alpha$. This eigenvalue problem is now solved using the same procedure as in the
previous section.

We first rewrite the system \rr{perturb2} and \rr{perturb2b} in the form \rr{compound1} but now $\xi$ is replaced by $Z$, the vector function ${\bm y}$  is given by  ${\bm y} = (\tilde{u}(Z),\tilde{u}'(Z),\tilde{w}(Z),\tilde{w}'(Z))^T$,
and the new coefficient matrix $A$  is not written out for brevity. We note, however, that $A$ is now a function of $Z$ via the fully nonlinear solution $(\bar{r},\bar{z})$, and also dependent on the value of $\rinf$. From the conditions governing the decay of the underlying state as $Z \to \pm \infty$ we require $\bar{r}(Z) \to \rinf,\; \bar{z}'(Z) \to \zinf$, and hence the matrix $A_\infty$ now takes the form

\begin{equation}
A_{\infty} = \left(
               \begin{array}{cccc}
                 0 & 1 & 0 & 0 \\
                 \frac{\omega(r_\infty)}{\hat{W}_{22}^{(\infty)}} & 0 & 0 & \frac{\hat{W}_1^{(\infty)} - \zinf \hat{W}_{12}^{(\infty)}}{\zinf \hat{W}_{22}^{(\infty)}} \\
                 0 & 0 & 0 & 1 \\
                 0 & \frac{-\hat{W}_1^{(\infty)} + \zinf \hat{W}_{12}^{(\infty)}}{ \hat{W}_{2}^{(\infty)}} & \frac{-\zinf \hat{W}_{1}^{(\infty)} + \zinf \rinf (\omega + \hat{W}_{11}^{(\infty)})}{\rinf \hat{W}_2^{(\infty)}} & 0 \\
               \end{array}
             \right). \label{Ainf}
\end{equation}
It is found again that the four eigenvalues of $A_{\infty}$ take the form  $\pm \hat{k}_1, \pm \hat{k}_2$. For $\rinf$ close to $r_{cr}$, $ \hat{k}_1$ and $ \hat{k}_2$ are real for $\alpha \in [0, \alpha_1]$, complex for $\alpha \in [\alpha_1, \alpha_2]$, and real again for $\alpha \in [\alpha_2, \infty)$ for
 some $\alpha_1$ and $\alpha_2$ that can only be determined numerically and are dependent on the value of $r_\infty$. Thus, at the isolated values  of $\alpha_1$ and $\alpha_2$, we have $ \hat{k}_1= \hat{k}_2$. These two isolated cases can be accommodated in the same way as $1/4$ is taken care of in the previous section.  As $\rinf$ moves away from $r_{cr}$, $\alpha_1$ and $\alpha_2$ coalesce and $ \hat{k}_1$ and $ \hat{k}_2$ are then real for all $\alpha$.

We proceed with the Evans function method outlined in the previous section to find the eigenvalues for each $\rinf$. Again, a single positive real eigenvalue is found for each $\rinf$. Close to the critical point, $r_{cr}$, we expect to recover the near-critical results discussed in Section 5. The connection between the eigenvalues $\sigma$ and $\eta$ is given by $\eta = \epsilon \sigma$. Hence in the limit as $\rinf \to r_{cr}$ we require the connection
\begin{equation}
\alpha =\frac{\rho \epsilon^2 \zeta_1}{c_1}, \label{weakfullconnection}
\end{equation}
where $\zeta_1$ is the only positive real eigenvalue found in Section 5, i.e. $\zeta_1 = 3/16$. This connection also provides the value of $\alpha_1$ where $\hat{k}_1=\hat{k}_2$, if $\zeta_1 $ is replaced by $1/4$.

\begin{figure}[!ht]
 \psfrag{aa}{$\rinf$}
 \psfrag{bb}{$\alpha$}
\centering \epsfig{figure=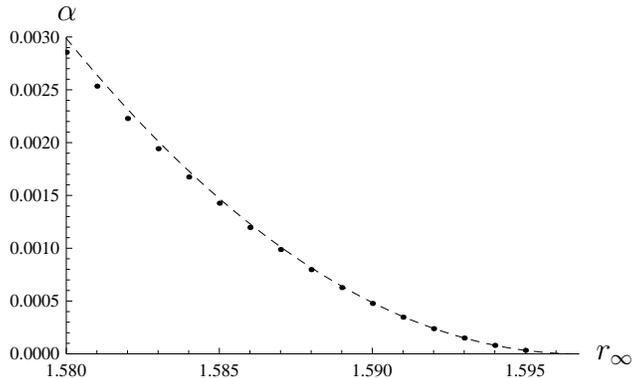,width=.5\textwidth}
\caption{Dependence of $\alpha$, defined by \rr{alpha}, on $\rinf$ for the closed-end Gent tube with $J_m=97.2$ and values of $\rinf$ in a small left neighborhood of the
first bifurcation value $1.59676$. Dotted line: numerical result based on
the fully nonlinear solution; dashed line: asymptotic result $\alpha =  \rho \zeta_1 \epsilon^2/c_1  = 10.6311 (\rinf-r_{cr})^2$ based on the weakly nonlinear solution.}
%
\label{genteigcompare}
\end{figure}

The above correspondence is confirmed in the limit as $\rinf \to r_{cr}$, for both the open and closed tubes described by the various strain-energy functions considered here. Figure \ref{genteigcompare} shows how the eigenvalue for the Gent  strain-energy function with closed ends is proportional to $(\rinf-r_{cr})^2$ near $r_{cr}$, with the coefficient given by $\rho \zeta_1/c_1$.

\begin{figure}[!ht]
 \psfrag{aa}{$\rinf$}
 \psfrag{bb}{$\alpha$}
\centering \epsfig{figure=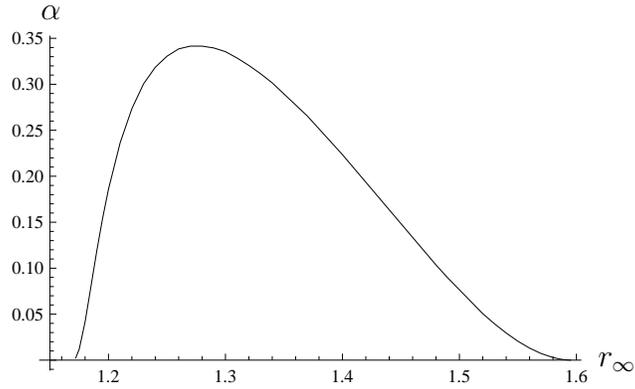,width=.5\textwidth}
\caption{Dependence of $\alpha$ on $r_\infty$ for the closed Gent tube with $J_m=97.2$, showing the fact that $\alpha$ tends zero as
the solitary-wave type solution tends to zero or the kink-wave type solution.}
\label{genteigclosed}
\end{figure}

%
%
%
%
Figure \ref{genteigclosed} shows how the single eigenvalue varies with respect to $r_\infty$ for a closed-end Gent tube with $J_m=97.2$. It is seen that as the kinked solution is approached at $r_k = 1.1694$, the value of the eigenvalue rapidly approaches zero. This is also the case when $J_m =30$ and $97.2$ with both closed and open ends, as well as the closed Ogden tube. This suggests that the kink-wave type solution is probably stable, although the possibility of other complex eigenvalues on the right half complex plane has not been
eliminated.

\begin{figure}[!ht]
 \psfrag{aa}{$\rinf$}
 \psfrag{bb}{$\alpha$}
\centering \epsfig{figure=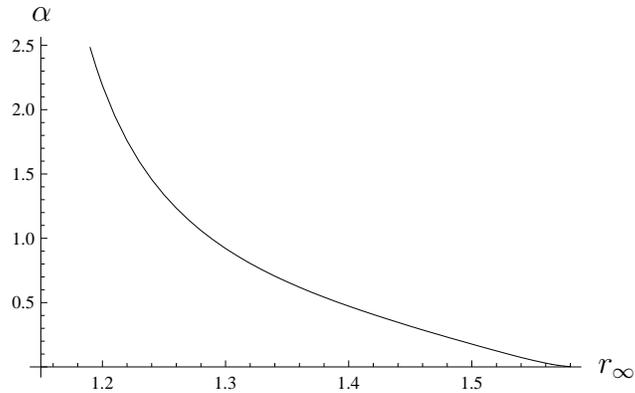,width=.5\textwidth}
\caption{Dependence of $\alpha$ on $r_\infty$ for the closed Varga tube.}
\label{vargaeigclosed}
\end{figure}

For the closed Varga tube we find that the value of the eigenvalue exponentially grows as $\rinf \to r^{*}$, where the eigenvalue tends to infinity, as can be shown in Figure \ref{vargaeigclosed}.

\section{Conclusion}
In this paper we have presented a graphical method for characterising the entire inflation process, and have studied the stability of the bifurcated solutions by determining whether there is a localised perturbation that would grow exponentially in time. The graphical method is based on the $r_0$ versus $\rinf$ diagram that gives almost all the information about the entire inflation and deflation process. Our spectral stability analysis shows that when the inflation is pressure controlled, the solitary-wave type solutions are unstable with respect to axi-symmetric perturbations. Our analysis seems to indicate that the kink-wave type solution might be stable, but as with all stability analysis, it is much
harder to establish stability than to prove instability. As remarked earlier, pressure-controlled inflation can be realized by connecting the inflating gas in the tube to a very large reservoir of the same gas, but unfortunately all available experimental results have been obtained for volume-controlled inflation only and so our theoretical predication is yet to be verified
by further experiments. We have also made some preliminary study on the stability of the bulging solutions using the energy method. The corresponding results will be reported in a separate paper which focuses on the case of volume-controlled inflation.

\subsection*{Acknowledgements}
This work is supported by a joint grant from BBSRC and EPSRC (BB/D014786/1)
under their Stem Cell Science and Engineering Initiative.

\section*{References}

\begin{enumerate}

\item Ablowitz, M.J. and Clarkson, P.A. (1991) {\it Solitons, nonlinear evolution equations and inverse scattering}. LMS Lecture Note Series 149,
Cambridge University Press.

\item Afendikov, A.L. and Bridges, T.J. (2001) Instability of the Hocking-Stewartson pulse and its implication for three-dimensional Poiseuille flow.
{\it Proc. R. Soc. Lond.} A {\bf 457}, 257-272.

\item Alexander, H. (1971) The Tensile Instability of an inflated cylindrical membrane as affected by an axial load. {\it Int. J. Mech. Sci.} {\bf 13}, 87-95.

\item Alexander, J.C., Gardner, R. and Jones, C.K.R.T. (1990) A topological invariant arising in the stability analysis of traveling waves. {\it J. Reine Angew. Math.}
{\bf 410}, 167-212.

\item Alexander, J.C. and Sachs, R. (1995) Linear instability of solitary waves of a Boussinesq-type equation: A computer assisted computation.
{\it Nonlin. World} {\bf 2}, 471-507.

\item Benedict, R., Wineman, A. and Yang, W.H. (1979) The determination of limiting pressure in simultaneous elongation and inflation of nonlinear elastic tubes. {\it Int. J. Solids Struct.} {\bf 15}, 241-249.

\item Bridges, T.J. (1999) The Orr-Sommerfeld equation on a manifold. {\it Proc. R. Soc. Lond.} A {\bf 455}, 3019-3040.

\item Budiansky, B. (1968) Notes on nonlinear shell theory. {\it J. Appl. Mech.} {\bf 35}, 393-401.

\item Chater, E. and  Hutchinson, J.W. (1984) On the propagation of bulges and buckles. {\it ASME J. Appl. Mech.}
{\bf 51}, 269-277.

\item Chen, Y-C. (1997) Stability and bifurcation of finite deformations
of elastic cylindrical membranes - part I. stability analysis. {\it Int.
J. Solids Struct.} {\bf 34}, 1735-1749.

\item Corneliussen, A.H. and  Shield, R.T.  (1961) Finite deformation of
elastic membranes with application to the stability of an inflated
and extended tube. {\it Arch. ration. Mech. Anal. } {\bf 7},
273-304.

\item Ericksen, J.L. (1975)  Equilibrium of bars.  {\it J. Elast.}
{\bf 5}, 191-201.

\item Evans, J.W. (1975) Nerve axon equations. IV. The stable and unstable impulse. {\it Indiana Univ. Math. J.} {\bf 24}, 1169-1190.

\item Epstein, M. and Johnston, C.R. (2001) On the exact speed and amplitude of solitary waves in fluid-filled elastic tubes. {\it Proc. R. Soc. Lond. A} {\bf 457}, 1195-1213.

\item Fu Y.B. and Il'ichev, A. (2009) Solitary waves in fluid-filled elastic tubes: existence, persistence, and the role of axial displacement. {\it IMA J. Appl. Math.}, submitted.

\item Fu, Y.B., Pearce, S.P. and Liu, K.K. (2008) Post-bifurcation analysis of a thin-walled hyperelastic tube under inflation. {\it Int. J. Non-linear Mech.} {\bf 43}, 697-706.

\item Fu, Y.B. and Zhang, Y.T. (2006) Continuum-mechanical modelling of kink-band formation in fibre-reinforced composites. {\it Int. J. Solids Strut.} {\bf 43}, 3306-3323.

\item Gilbert, F. and Backus, G.E. (1966) Propagation matrices in elastic wave and vibration problems. {\it Geophysics} {\bf 31}, 326-332.

\item Goncalves, P. B., Pamplona, D. and Lopes, S.R.X (2008) Finite deformations of an initially stressed cylindrical shell under internal pressure. {\it Int. J. Mech. Sci.} {\bf 50}, 92-103.

\item Gent, A. N. (1996)  A new constitutive relation for
rubber. {\it Rubber Chem. Technol.} {\bf 69}, 59-61.

\item Haughton, D.M. and Merodio, J. (2009) The elasticity of arterial tissue affected by Marfan¡¯s syndrome.
{\it Mech. Res. Comm.} {\bf 36}, 659-668.

\item Haughton, D.M. and Ogden, R.W. (1979)  Bifurcation of
inflated circular cylinders of elastic material under axial loading. I.
Membrane theory for thin-walled tubes. {\it J. Mech. Phys. Solids} {\bf
27}, 179-212.

\item Horgan, C.O. and G. Saccomandi, G. (2003) A description of
arterial wall mechanics using limiting chain extensibility
constitutive models. {\it Biomech. Model.
Mechanobiol.} {\bf 1},  251-266.

\item Humphrey, J.D. and Canham, P.B. (2000) Structure, mechanical properties, and mechanics of intracranial saccular aneurysms. {\it J. Elast.} {\bf 61}, 49-81.

\item Hutchinson and Neale, K.W. (1983) Neck propagation. {\it J. Mech. Phys. Solids} {\bf 31}, 405-426.

\item Kanner, L.M. and Horgan, C.O. (2007) Elastic instabilities for strain-stiffening rubber-like spherical and cylindrical thin shells under inflation {\it Int. J. Non-linear Mech.} {\bf 42}, 204-215.

\item Kydoniefs, A.D. and Spencer, A.J.M. (1969) Finite
axisymmetric deformations of an initially cylindrical elastic membrane.
{\it Q. Jl. Mech. Appl. Math.} {\bf 22}, 87-97.

\item Kyriakides, S. (1981) On the stability of inelastic circular pipes under combined bending and external pressure. {\it Proceedings of the 1981 SESA Spring Meeting} 372-378.

\item Kyriakides, S. and Chang, Y.-C. (1990) On the initiation of a long elastic tube in the presence of axial load. {\it
Int. J. Solids Struct.} {\bf 26}, 975-991.

\item Kyriakides, S. and Chang, Y.-C. (1991) The initiation and
propagation of a localized instability in an inflated elastic tube. {\it
Int. J. Solids Struct.} {\bf 27}, 1085-1111.

\item Ng, B.S. and Reid, W.H. (1979) An initial-value method for eigenvalue problems using compound matrices. {\it J. Comp. Phys.} {\bf 30}, 125-136.

\item Ng, B.S. and Reid, W.H. (1985) The compound matrix method for ordinary differential systems. {\it J. Comp. Phys.} {\bf 58}, 209-228.

\item Ogden, R.W. (1972) Large deformation isotropic elasticity-on the correlation of theory and experiment for
incompressible rubber-like solids. {\it Proc. R. Soc. Lond.} A {\bf 326}, 565-584.

\item Ogden, R.W. (1997) {\it Non-linear elastic deformations}. New York: Dover.

\item Pamplona, D.C., Goncalves, P.B. and Lopes, S.R.X. (2006)
Finite deformations of cylindrical membrane under internal pressure. {\it
Int. J. Mech. Sci.} {\bf 48}, 683-696.

\item Pipkin, A.C. (1968) Integration of an equation in membranes
theory. {\it ZAMP} {\bf 19}, 818-819.

\item Shi, J. and Moita, G.F. (1996) The post-critical analysis of
axisymmetric hyper-elastic membranes by the finite element method. {\it
Comput. Methods Appl. Mech. Engrg.} {\bf 135}, 265-281.

\item Shield, R.T. (1972)  On the stability of finitely deformed
elastic membranes; Part II: Stability of inflated cylindrical and
spherical membranes. {\it ZAMP}  {\bf 23}, 16-34.

\item Vorp, D.A. (2007) Biomechanics of abdominal aortic aneurysm. {\it J. Biomechanics} {\bf 40}, 1887-1902.

\item Watton, P.N., Hill, N.A. and Heil, M. (2004) A mathematical model for the growth of the abdominal aortic aneurysm.
{\it Biomechan Model Mechanobiol} {\bf 3}, 98-113.

\item Yin, W.-L. (1977) Non-uniform inflation of a cylindrical
elastic membrane and direct determination of the strain energy function.
{\it J. Elast.} {\bf 7}, 265-282.

\end{enumerate}

\end{document}